\documentstyle[11pt,newpasp,twoside,epsf]{article}
\def\edcomment#1{\iffalse\marginpar{\raggedright\sl#1\/}\else\relax\fi}
\reversemarginpar

\begin{document}
\title{Tracing the Cluster Internal Dynamics with Member Galaxies}
 \author{Andrea Biviano}
\affil{INAF -- Osservatorio Astronomico di Trieste, 
via G.B. Tiepolo 11, I-34131 Trieste, Italy -- biviano@ts.astro.it}

\begin{abstract}
The analysis of the spatial distribution and kinematics of galaxies in
clusters allows one to determine the cluster internal
dynamics. In this paper, I review the state of the art of this topic. 
In particular, I summarize what we have learned so far
about galaxy orbits in clusters, and about the cluster mass
distribution. I then compare four methods that have recently been used
in the literature, by applying them to the same data-set. The results
stress the importance of reducing {\em systematic} besides {\em
random} errors, for a reliable determination of cluster mass
profiles. 
\end{abstract}

\section{Introduction}
Already in 1931, Hubble \& Humason (1931) provided rough estimates of
the velocity dispersions of three galaxy clusters, ranging from 500 to
1200 km~s$^{-1}$. This is more or less the range of values found for
rich clusters nowadays. A few years later, Zwicky
(1933, 1937) and Smith (1936), determined the total masses of the Coma
and the Virgo cluster, respectively, by application of the virial
theorem to the cluster galaxies.  The cluster masses they
determined\footnote{Throughout this paper I use a Hubble constant
H$_0$=70~km~s$^{-1}$~Mpc$^{-1}$. Quoted values are scaled accordingly.},
$> 3.3 \times 10^{14} M_{\sun}$ (Coma), and
$7.7 \times 10^{14} M_{\sun}$ (Virgo), are in reasonable good
agreement with nowadays estimates.
Of course, these early estimates used
galaxies as unbiased tracers of the underlying mass distribution, but
this was just a convenient assumption.  Relaxing this assumption can
have a large effect on the cluster mass estimate (see, e.g., Merritt
1987). In fact, galaxies are {\em unlikely} to be distributed like the
mass, simply because different cluster galaxy populations have
different distributions (e.g. emission-line galaxies, ELG hereafter,
and non-ELG, see Biviano et al. 1997). That is why we need to
determine the cluster mass distribution, if we want to determine its
total mass. And if we use the cluster galaxies for this purpose, we
must at the same time determine their orbital anisotropies.

In this contribution, I review our current understanding of the
internal dynamics of clusters, as determined from the spatial and
velocity distributions of cluster member galaxies. Due to space
limitation, I focus my review on the determination of the cluster
mass, and mass-to-light profiles, and on the orbital anisotropy of
cluster galaxies. Readers interested in cluster substructures are
referred to the recent review of Girardi \& Biviano (2001), and those
interested in a complete historical review of this topic are referred
to Biviano (2000).

\section{The Jeans approach}
Here I briefly describe the traditional Jeans approach (see Binney
\& Tremaine 1987).  For the sake of simplicity, I assume to deal with
a non-rotating spherically symmetric system, in a steady-state.  The
Abel equations relate the projected density profile, $I(R)$, and
projected velocity dispersion profile, $\sigma_p(R)$ (the observables),
to the de-projected density profile, $\nu(r)$ and, respectively, the
radial and tangential components of the velocity dispersion profile,
$\overline{v_r^2}$, and $\overline{v_t^2}$. $R$ and $r$ are the
projected and 3D clustercentric radial distances, respectively.  
When $\nu(r)$, $\overline{v_r^2}$ and $\beta(r)$ are determined,
the mass profile follows from the Jeans equation.

In principle, one should determine the de-projected profiles from the
observed ones, but this is not simple, unless the anisotropy
term, $\beta(r) \equiv 1-\overline{v_t^2}/\overline{v_r^2}$ is zero. The
general case requires solving an integro-diffential equation (eq.4-62
in Binney \& Tremaine 1987; see Binney \& Mamon 1982; Solanes \&
Salvador-Sol\'e 1990). This ``direct'' approach is
complicated when the data are noisy. Therefore, the simpler
``inverse'' route is usually chosen. One adopts suitable functional
forms for the 3D profiles -- $\nu(r)$, $\overline{v_r^2}$, $\beta(r)$,
or the mass profile, $\rho(r)$ --, projects them in 2D, and
determines their best-fit parameters by comparison with the
observables (see, e.g., Carlberg et al. 1997a, and Sect.~3.2).

The Jeans approach has several potential problems. First of all, one
generally assumes the system to be spherically symmetric, but clusters
are not. However, the number of available galaxies in each single
cluster is limited, and it is often necessary to build a ``composite''
cluster by joining together the data for several clusters, after
proper normalization (see, e.g., Biviano et al.  1992, 1997; Carlberg
et al. 1997a). This composite cluster is spherically symmetric by
construction.

Evidence for cluster rotation has been claimed for a couple of
clusters (see, e.g., Biviano et al. 1996; Dupke \& Bregman 2001) but
in general we can safely assume that rotational motions are not
important.

The assumption of steady-state is not always valid. E.g. galaxy
groups are probably still in the phase of collapse (Giuricin et
al. 1988). On the other hand, in a low-density Universe, clusters
evolve quite rapidly, and their formation process should be over by
now.  Following Ellingson et al. (2001), we estimate that the infall
rate of field galaxies in nearby clusters is $\la 1$\% of the total
cluster population per Gyr. Time-dependent terms in the Jeans equation
could be needed in more distant clusters, or when galaxies are subject
to tidal disruption (e.g. spirals in cluster centres, see Katgert
et al. 2002).

The Jeans equation is derived from the collisionless Boltzmann
equation, and its applicability is limited to collisionless
fluids. Clusters of galaxies are not collisionless in their inner
regions (e.g. Biviano et al. 1992; den Hartog \& Katgert 1996), and
dynamical friction and mergers could be important for very massive
custer galaxies. It is then wise to exclude the brightest cluster
members from the sample (Katgert et al. 2002).

Interlopers are not a serious problem, when one deals with a composite
sample of several hundred galaxies. Different methods of interloper
rejection give similar results (Fadda et
al. 1996). Subclustering is a more serious issue, since galaxies in
subclusters are distributed in a different way from the smooth cluster
population (Biviano et al. 2001; see also Katgert's contribution in
these proceedings). It is better to identify galaxies belonging to
subclusters, and to exclude them from the sample (Katgert et al.
2002).

Finally, the basic problem of the ``Jeans'' approach is that the
solution is intrinsically degenerate; we can either assume $\beta(r)$
and determine $M(<r)$, or {\em viceversa,} but we cannot simultaneously
constrain $M(<r)$ and $\beta(r)$ from the observables (see, e.g., King
1972; Bailey 1982). A possible solution consists in modelling the
whole velocity distribution of cluster galaxies, and not only its
second moment (Merritt 1987; Merritt \& Saha 1993). Another
possibility is to break this degeneracy by the use of several
different tracers of the same gravitational potential (Katgert et
al. 2002; see Sect.~3.6).

\section{Mass profiles and orbital anisotropies: results so far}
Seven decades since its first application to Coma (Zwicky 1933) the
virial theorem has not found simple alternatives yet (see Perea et
al. 1990 and references therein). The main improvement over the
classical virial method has been the introduction of the
``surface-term'' (The \& White 1986; see also Sect.~3.4). The
solutions to the direct Jeans approach provided by Binney \& Mamon
(1982) and Solanes \& Salvador-Sol\'e (1990), have never been applied
to galaxy cluster data-sets so far, because these
data are noisy. Most analyses therefore model the profiles
instead of deriving them directly from the data.  If the models are
not general enough, this procedure can lead to biased results. That is
why non-parametric approaches have been developed, mostly by Merritt
and collaborators (Merritt \& Saha 1993; Merritt \& Tremblay 1994;
Merritt \& Gebhardt 1994; Gebhardt \& Fisher 1995).

Despite these technical improvements, the question of whether or not
the cluster mass is distributed like the galaxies has not received a
definite answer yet. In clusters, as well as in groups, recent works
provide evidence both in favour (Carlberg et al. 1997a; Mahdavi
et al. 1999; van der Marel et al. 2000), and against (Koranyi et
al. 1998; Rines et al. 2000; Carlberg et al. 2001) the hypothesis of a
constant mass-to-light ratio with radius.  Cluster mass density
profiles are found to be in agreement with the 'universal'
profile of Navarro, Frenk \& White (1997; NFW hereafter), but other
models are acceptable as well (Carlberg et al. 1997b; van der Marel et
al. 2000; Geller, Diaferio \& Kurtz 1999). From a theoretical point of
view, the situation is not much clearer.  Both decreasing and
increasing mass-to-light profiles are predicted, resulting from
the processes of tidal stripping (Mamon 1999), or, respectively, of
dynamical friction and merging (Fusco-Femiano \& Menci 1998).

Numerical simulations show that dark matter particles
are characterized by a moderate radial anisotropy, increasing out to
$\sim 2 \, r_{200}$ (e.g. Diaferio 1999). However, the idea of shells
of collapsing material around clusters has been around since the work
of Gunn \& Gott (1972). This work stimulated several investigations,
and the evidence for the infall of spirals into clusters has
been accumulating over the years. Indirect evidence comes from the larger
velocity dispersion of spirals as compared to that of early-type
galaxies (Moss \& Dickens 1977; Sodr\'e et al. 1989). More direct
evidence is provided by the analyses of the Tully-Fisher
distance-velocity diagram in nearby clusters (Tully \& Shaya 1984;
Gavazzi et al. 1991).

Additional constraints on the orbital anisotropy of cluster galaxies
come from the analysis of their velocity distributions. Most studies
conclude that the orbits of early-type galaxies are
quasi-isotropic, and those of late-type galaxies moderately radial
(Mohr et al. 1996; Biviano et al. 1997; Adami et al. 1998; van der
Marel et al. 2000, see also Sect.~3.3), with the exception of
Ram\'{\i}rez \& de Souza (1998, see Sect.~3.1). Similar results are
also found by the application of dynamical models (e.g. Natarajan
\& Kneib 1996; Carlberg et al. 1997c; Mahdavi et al. 1999).

Other methods for constraining the galaxy orbits include Pryor \&
Geller's (1984) and Merrifield's (1998). Pryor \& Geller try to constrain
the orbits of HI-deficient galaxies by noting that cluster core-crossing
is a necessary condition for gas stripping. Merrifield suggests
using the radio or X-ray trails of galaxies in order to constrain their
orbits.

\subsection{Ram\'{\i}rez \& de Souza}
Ram\'{\i}rez \& de Souza (1998) propose to use $\mid u \mid
\equiv <\mid v_j \mid>/\sigma_p$ as a robust shape estimator of the
velocity distribution of cluster members. The value of the
anisotropy of galaxy orbits is obtained by comparing the observed
values of $\mid u \mid$ with those derived from a kinematical model. The
model is rather over-simplified, and not very realistic: the velocity
distribution is Gaussian at any radial distance, and both
$\overline{v_r^2}$ and $\overline{v_t^2}$ are constant. 

Van der Marel et al.  (2000) point out that the assumption of
Gaussianity is not acceptable for anisotropic models. Another problem
is that Ram\'{\i}rez \& de Souza (1998) do not compute $\sigma_p$ and
$<\mid v_j \mid>$ on the same sample.  E.g., when they consider the
subsample of ellipticals, $\sigma_p$ is still computed on all cluster
galaxies. This effectively makes the $\mid u \mid$ parameter a {\em
scale} rather than a {\em shape} estimator. They find that ellipticals
have a lower $\mid u \mid$ than spirals, but this is just
a restatement of the finding of Moss \& Dickens (1977; see also Adami
et al. 1998), that ellipticals have a lower velocity dispersion than
spirals, and do not support Ram\'{\i}rez \& de Souza's (1998) claim of
radial anisotropy.

\subsection{Carlberg et al.}
Carlberg et al. (1997a,b) analyse the
internal dynamics of galaxy clusters from CNOC, by the ``inverse''
Jeans approach (see Sect.~2). They adopt {\em ad hoc} functional
forms for $\nu(r)$, $\overline{v_r^2}$, and $\beta(r)$, project them
in 2D, and constrain their parameters via a $\chi^2$ comparison with
the observables. The advantage of their approach is that the chosen
functions are well behaved, and easy to invert, but the solution for
$M(<r)$ is to some extent imposed {\em a priori} (Merritt \& Gebhardt
1994).

Their results are based on $\sim 1000$ galaxy members in a
composite of 14 ``regular'' clusters. They find that galaxies
trace the mass, to within $\pm 30$\%, and the anisotropy is null or at
most mildly radial, $0 \la \beta \la 0.5$. The mass profile is found
to be consistent with a NFW profile, but other profiles (Hernquist
1990) provide equally acceptable, if not better, fits.

Carlberg et al. (2001) have recently applied the same technique on a
composite sample of $\sim 700$ galaxies in $\sim 200$ groups from the
CNOC2. They find a strongly increasing mass-to-light ratio with
radius, at variance with the findings of Mahdavi et al. (1999). The
construction of a composite group sample is anyway problematic, due to
uncertainties in the mean velocity and velocity dispersions of
individual groups. Moreover, removing interlopers in small galaxy
groups is a difficult task. Finally, groups may be far from
virialization (Giuricin et al. 1988).  Therefore, I feel that Carlberg
et al.'s result needs confirmation by independent analyses.

\subsection{van der Marel et al.}
The CNOC sample has recently been re-analysed by van der Marel et
al. (2000) using the method developed by van der Marel (1994). They
assume a three-parameter family of models for the mass density
profile, $\rho(r)$, and a set of values for the anisotropy $\beta$,
taken to be a constant over radii, and determine the model parameters
from the best-fit to $\sigma_p(R)$.  The galaxy density profile is
derived via Abel-inversion from a parametric representation of
$I(R)$. Their approach improves over Carlberg et al.'s, since both the
adopted model for $\rho(r)$, and the function representing $I(R)$, are
quite general.  On the other hand, a constant $\beta$ is unlikely to
be a realistic model.

In order to break the intrinsic degeneracy between $\rho(r)$ and
$\beta$, van der Marel et al. take into account the shape of the whole
velocity distribution of cluster members (following the idea of
Merritt \& Saha 1993). They are thus able to conclude that mass
follows light to within $\pm 30$\%, $\beta$ ranges from $-0.8$ to 0.1,
and a NFW profile is an acceptable fit to $\rho(r)$ -- but not unique.

\subsection{Girardi et al.}
Girardi et al. (1998) work out the masses of 170 clusters, the
largest sample analysed so far, using the virial theorem with the
surface-term correction. In order to work out this correction,
they first determine $\nu(r)$ from the Abel inversion of a (simple)
parametric representation of $I(R)$. Assuming $\rho(r) \propto
\nu(r)$, they then determine which of three $\beta(r)$-models
provides the best-fit to $\sigma_p(R)$. From $\beta(r)$ and $\rho(r)$,
$\overline{v_r^2}(r)$ and the surface-term correction are derived.
In principle, it is then possible to compute the corrected virial
mass at any radius, hence the mass profile, but the solution is
certainly biased by the rather arbitrary assumptions.

They find that the surface-term correction is important, of order
20\% at the virial radius (in agreement with The \& White 1986, and
Carlberg et al. 1997b). They find good agreement between their
optically-determined cluster masses and those derived from the X-ray
data. Using the same algorithm, Girardi \& Mezzetti (2001) claim no
evidence for the evolution of the internal cluster dynamics up to $z
\simeq 0.3$.

\subsection{Diaferio \& Geller}
Based on the results of numerical simulations of structure formation,
Diaferio \& Geller (1997) and Diaferio (1999) have
devised a new method for determining cluster mass profiles. They
show that the velocity field of cluster surroundings is determined
by the cluster mass distribution, via $GM(<r)=\int_0^r {\cal A}^2(x)
\, {\cal F}_{\beta}(x) \, dx$ where ${\cal A}$ is the amplitude of the
caustics in the $(R,v)$ space, and ${\cal F}_{\beta}$ is a function of
the gravitational potential and of the anisotropy profile, assumed to
be a constant.  The caustics are determined by an adaptive kernel
method.

The ``caustic method'' does not rely upon the hypothesis of virial
equilibrium, and thus it allows estimating $M(<r)$ to large radii,
where other methods fail. On the other hand, the choice of the caustic
in the $(R,v)$ plane is rather subjective, and the approximation
${\cal F}_{\beta} \simeq$~constant is valid only beyond
1--2~$r_{200}$.

Using this method Geller et al. (1999) estimate $M(<r)$
of the Coma cluster out to 14~Mpc, and find it to be well fitted by a
NFW profile. Rines et al. (2000) determine the mass-to-light profile
of Abell~576 out to 6~Mpc, and find it to be decreasing with radius.
A similar indication for little mass outside the virialized cluster
region is found by Reisenegger et al. (2000) in the Shapley
supercluster. However, the method can constrain $M(<r)$ only to within
a factor two (Diaferio 1999), and a constant mass-to-light profile is
probably still consistent with the data.

\begin{figure}
\plotfiddle{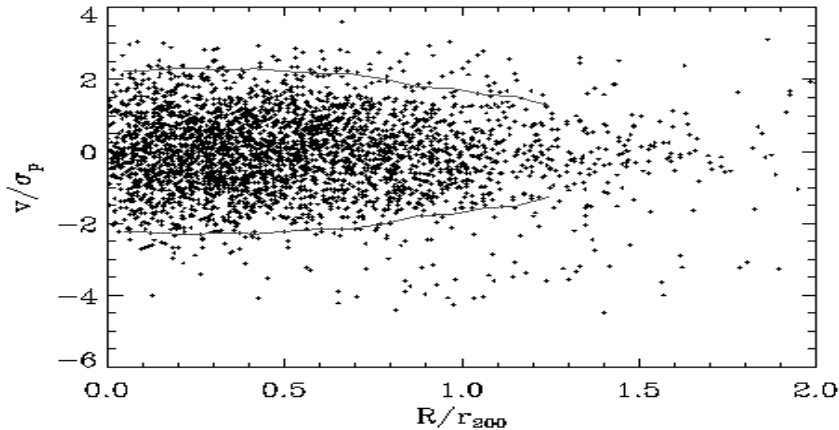}{5cm}{0}{70}{50}{-220}{-120}
\caption{The $(R,v)$ plane of 3365 galaxies in the composite
cluster sample drawn from ENACS. Solid lines show the location of
the caustics, as determined with the method of Diaferio (1999).}
\end{figure}
\subsection{Katgert \& Biviano}
Peter Katgert and myself, we are currently analysing the internal
dynamics of a composite data-set of $\sim 3000$ galaxies in clusters
from ENACS.  We build $M(<r)$ directly from $I(R)$ and $\sigma_p(R)$,
instead of modelling $\rho(r)$ and $\beta(r)$ with {\em ad hoc} chosen
functions. In fact, {\em ``using ad hoc parametrized models severely
bias the answer regardless of the quality of the data''} (Merritt \&
Gebhardt 1994).  Our approach is fully non-parametric; we use {\em
LOWESS} (Gebhardt \& Fisher 1995) to provide a smoothed representation
of the observed profiles, which we invert to provide the mass profile
in the isotropic case. Using several tracers of the gravitational
potential (i.e. ellipticals and S0's, early spirals, late spirals and
ELG) we expect to break the intrinsic degeneracy of the problem,
to constrain the orbits of cluster galaxies, and to test whether
different galaxy populations are in equilibrium within the same
gravitational potential.

The main problem of this method is numerical instability. This
can be solved by smoothing the data, but excessive smoothing
must be avoided, not to erase useful information in the observed
profiles.

Preliminary results (Biviano et al. 1999; Mazure et al. 2000) indicate
that different cluster galaxy populations are characterized by
different anisotropy profiles. 

\begin{figure}
\plottwo{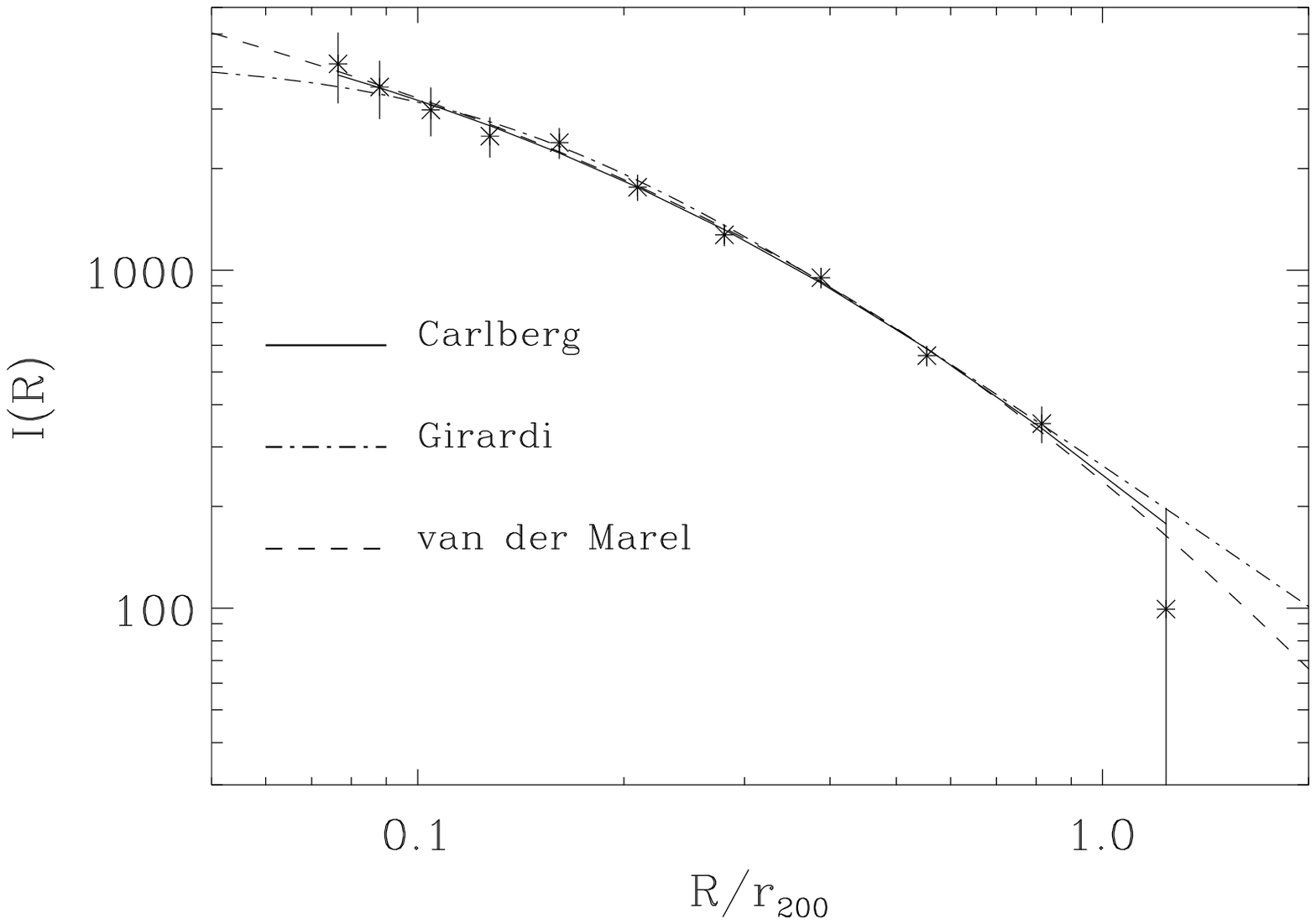}{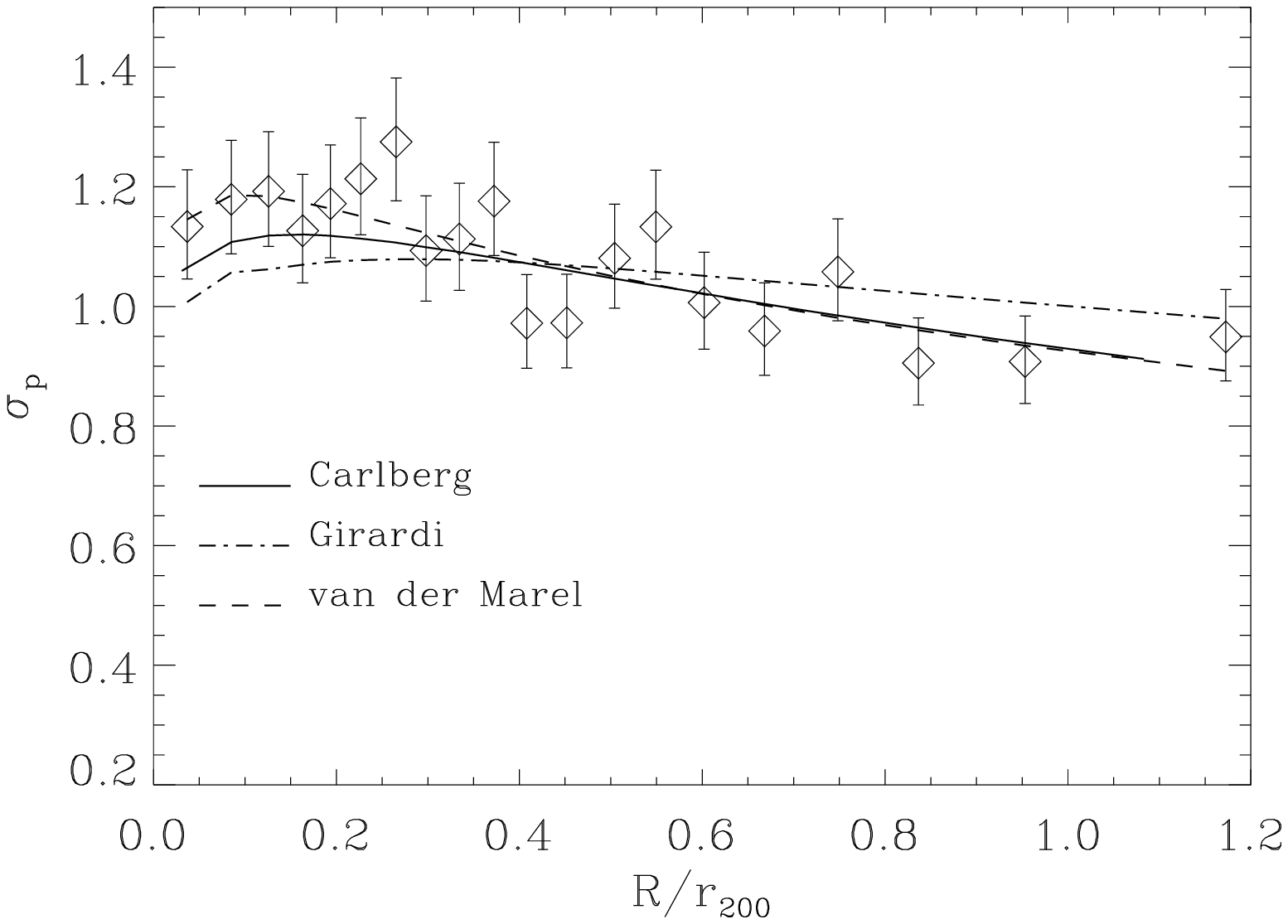}
\caption{The projected number density profile $I(R)$ (left panel) and
the projected velocity dispersion profile $\sigma_p(R)$ (right panel)
of the 1557 non-ELG members of the composite ENACS cluster. The
labelled lines indicate model fits of these profiles.}
\end{figure}
\section{Four new methods and one data-set}
In this Section I compare the results obtained by
four different methods on the {\em same} data-set.  I consider 37 ENACS
(Katgert et al. 1996, 1998) 
clusters with at least 45 galaxy redshifts, totalling 3365 galaxies.
The observational data are shown in Fig.~1, where I also display the
location of the caustics, determined with the method of Diaferio
(1999).  From these clusters I remove interlopers (following den
Hartog \& Katgert 1996), and galaxies with a significant probability
of being in subclusters (following Biviano et al. 2001). I then build
a composite cluster, in which galaxy velocities are scaled with
the cluster $\sigma_p$'s, and galaxy clustercentric distances are
scaled with the cluster $r_{200}$'s. There are 1557 non-ELG and 217
ELG in this ``cleaned'' composite cluster.  I apply the methods of Carlberg et
al., Girardi et al., and van der Marel et al. to
the sample of 1557 non-ELG. The last method is also
applied to the sample of 217 ELG. Diaferio \& Geller's metod is
instead applied on the uncleaned sample of 3365 galaxies, since it
does not require preliminary removal of interlopers, or subclusters.
Results of the method of Katgert \& Biviano will be presented
elsewhere (Katgert et al. 2002).

The advantage of using a composite cluster is that one reduces
asymmetries in the galaxy distributions, thus effectively imposing the
circular symmetry to the problem. But of course, the composite
cluster is an artificial construction, and as such it may bear no
resemblance to any real cluster. On the other hand, we may reasonably
expect that results obtained for this composite cluster are
representative of the average cluster population. It is important to
use a homogeneous sample (like ENACS, or CNOC) for properly dealing
with observational selection functions.  This is particulary important
for the determination of $I(R)$ (for further details, see Biviano et
al. 2001).

In Fig.~2a,b I plot the $I(R)$ and $\sigma_p(R)$ for the 1557 non-ELG.
Also shown are the three fits to these profiles,
determined using the three different methods labelled in the
Figure, and assuming $\beta(r) \equiv 0$.  The
most remarkable differences occur where the data are less
constraining, i.e. at small and large radii. They are caused by
different choices of the fitting functions.

\begin{figure}
\plotfiddle{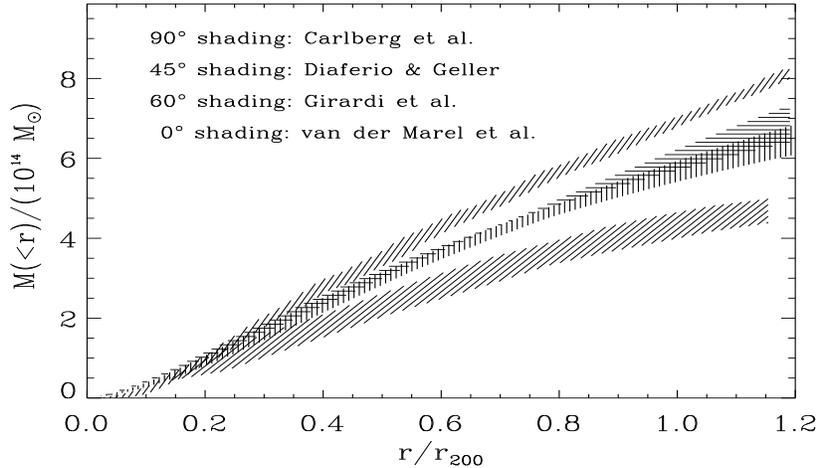}{5cm}{0}{70}{55}{-220}{-215}
\caption{Four different estimates of the mass profile of the
non-ELG composite cluster, under the isotropic assumption.
Shadings represent (approximate) 1-$\sigma$ uncertainties.}
\end{figure}
The mass profiles determined with the four different methods (assuming
$\beta(r) \equiv 0$ where needed) are shown in Fig.~3. The mass scale
is arbitrary and the different values of $M(<r_{200})$ indicate that
different effective values of $r_{200}$ are implied by the different
methods. Approximate 1-$\sigma$ confidence intervals are determined by
splitting the sample in two, and independently computing $M(<r)$ on
each subsample. The four different estimates of the mass profile are
not in agreement.  the mass at any given radius is uncertain by a
factor two, and systematic errors dominate over random errors.

In Fig.s~4a,b I show the integrated mass-to-number and
mass-to-luminosity ratios as a function of radius, computed with the
four different methods. On average, light seems to trace mass to
within $\pm 30$~\%, at radii $>0.4 \, r_{200}$. In the central region,
the mass seems to be more concentrated than the galaxies, but less
than their light, an indication of luminosity segregation (see, e.g.,
Biviano et al. 1992). 

The determination of the mass for the non-ELG is based on the
assumption of isotropy. This assumption is supported by the analysis
of the non-ELG velocity distribution, shown in Fig.~5a, alongwith its
Gauss-Hermite representation. The Gauss-Hermite moments of this
distribution clearly indicate that isotropy (or, at most, mild
tangential anisotropy, $\overline{v_r^2}/\overline{v_t^2} \simeq 0.8$)
is a valid assumption.

I then consider the sample of 217 ELG. I assume that the non-ELG mass
profile, in the isotropic assumption, is the true cluster mass
profile, and solve for $\beta(r)$ of the ELG, using the method of van
der Marel et al.  I show in Fig.~5b the best-fit $\chi^2$ solution for
a constant $\overline{v_r^2}/\overline{v_t^2}$. For consistency, the
solutions for non-ELG are also shown, where deviations from the
expected $\overline{v_r^2}/\overline{v_t^2} \equiv 1$ solution are
indicative of the accuracy of the method.  The data indicate a
moderate degree of radial anisotropy for the ELG, $1.3 \leq
\overline{v_r^2}/\overline{v_t^2} \leq 1.6$, at 68\% confidence
level. Zero anisotropy models for the ELG are rejected at $>99.99$\% probability.

\begin{figure}
\plottwo{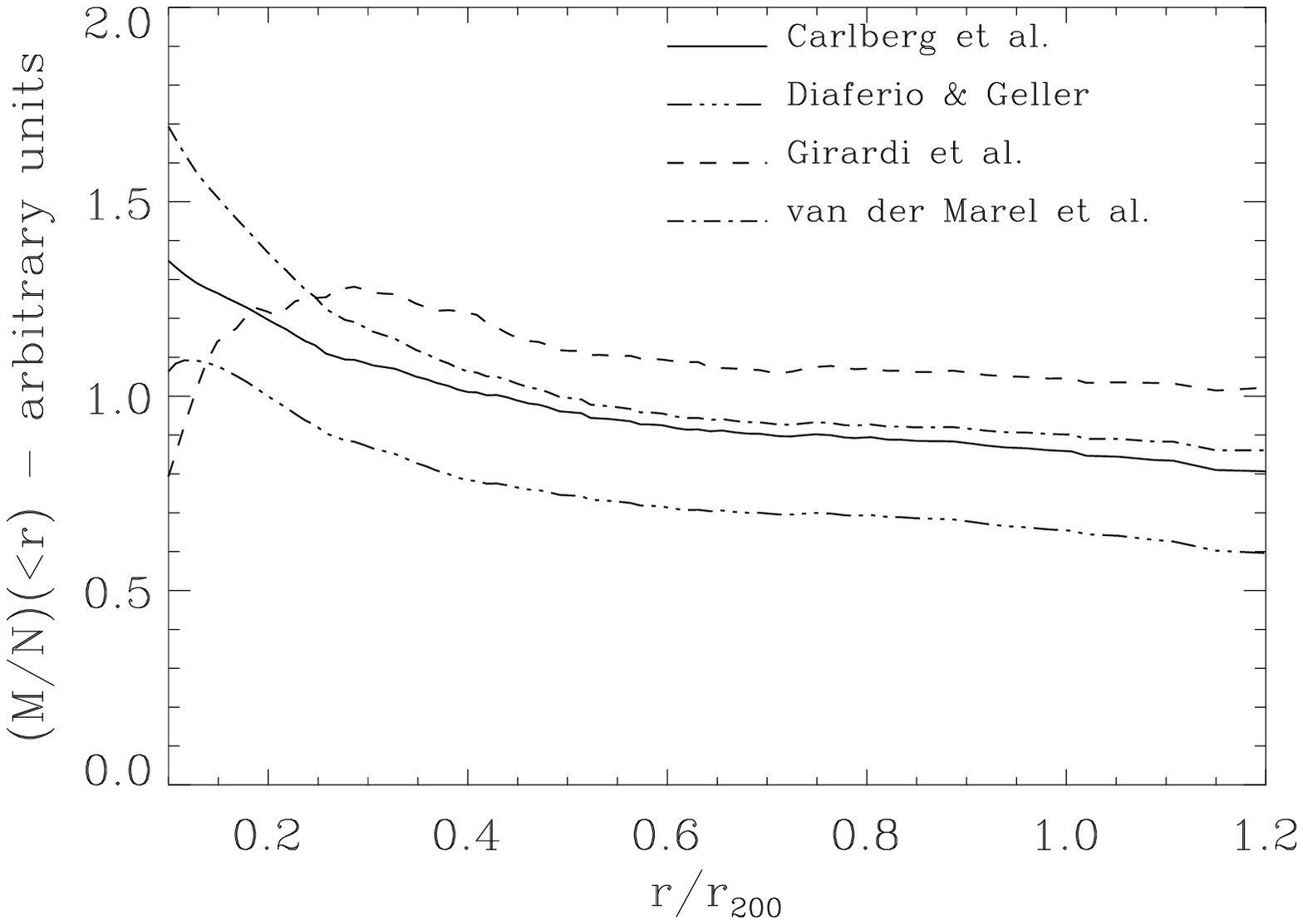}{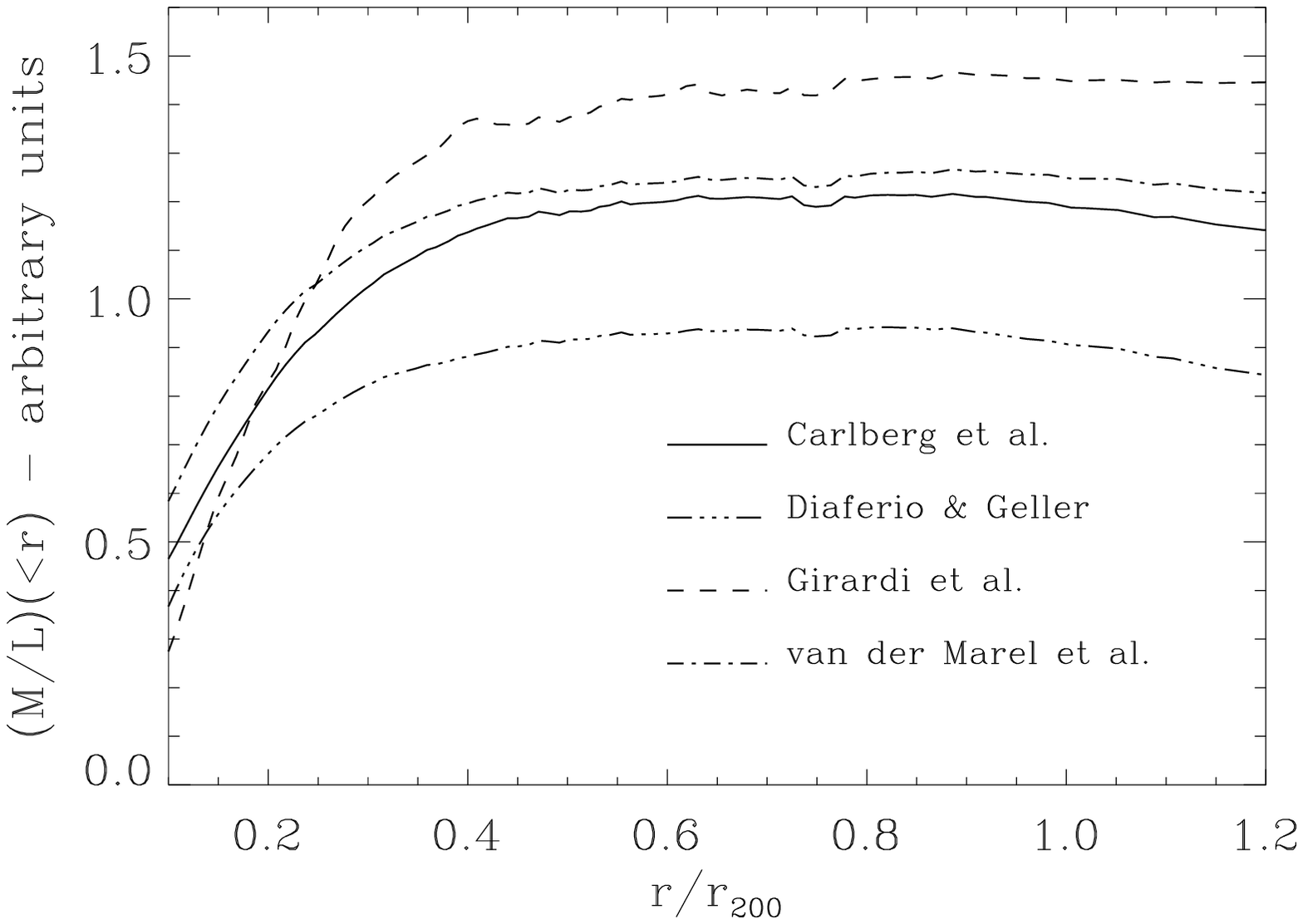}
\caption{Left panel: the integrated mass-to-number profile determined
with four different methods on the 
the 1557 non-ELG members of the composite ENACS cluster,
assuming isotropy.
Right panel: the integrated mass-to-light profile determined
with four different methods on the 
the 1557 non-ELG members of the composite ENACS cluster,
assuming isotropy.}
\end{figure}
\section{Conclusions}
The use of member galaxies to determine the internal cluster dynamics
is now reaching an unprecedented level of accuracy, thanks to the
increasing amount of data (ENACS, Katgert et al. 1996, 1998; CNOC, Yee,
Ellingson, \& Carlberg 1996; Coma, Geller et al. 1999; Abell 576,
Rines et al. 2000), and the development of new algorithms.  The size
of the data-sets must still be increased if we want to separately consider
different cluster galaxy populations (e.g.  early-types and
late-types).

Whatever the sample size, it is difficult to discriminate among
different models for the cluster mass density profile. This is because
models either differ at large radii, where the cluster fades into the
field, or near the centre, where one samples the cD rather than the
cluster potential.  The determination of cluster mass profiles appears
to suffer from the systematics of different methods.  Even on samples
of $\sim 1000$ galaxies, the total mass of a cluster can be uncertain
by a factor two. Such an uncertainty also affects our current
knowledge of the relative distributions of mass and light. Taking the
data at face value, there are now some indications that mass {\em is}
indeed {\em roughly} traced by light, at least outside the very central
region (where there is evidence for luminosity segregation). Finally,
independent analyses suggest that different cluster galaxy populations
are characterized by different orbits. The currently best-buy model
has (spectral or morphological) early-type galaxies on fairly
isotropic orbits, and late-type galaxies on moderately radial orbits.

\begin{figure}
\plottwo{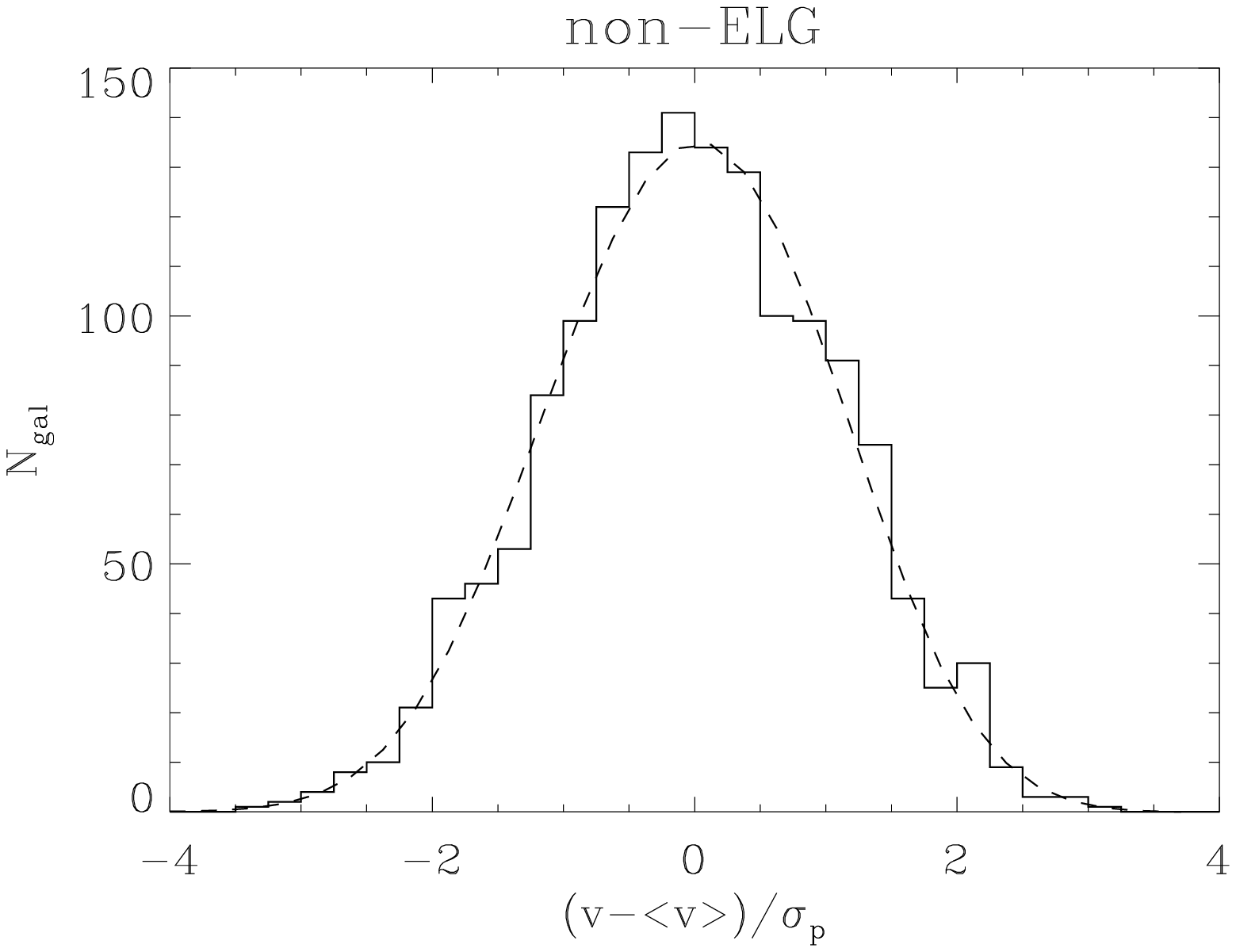}{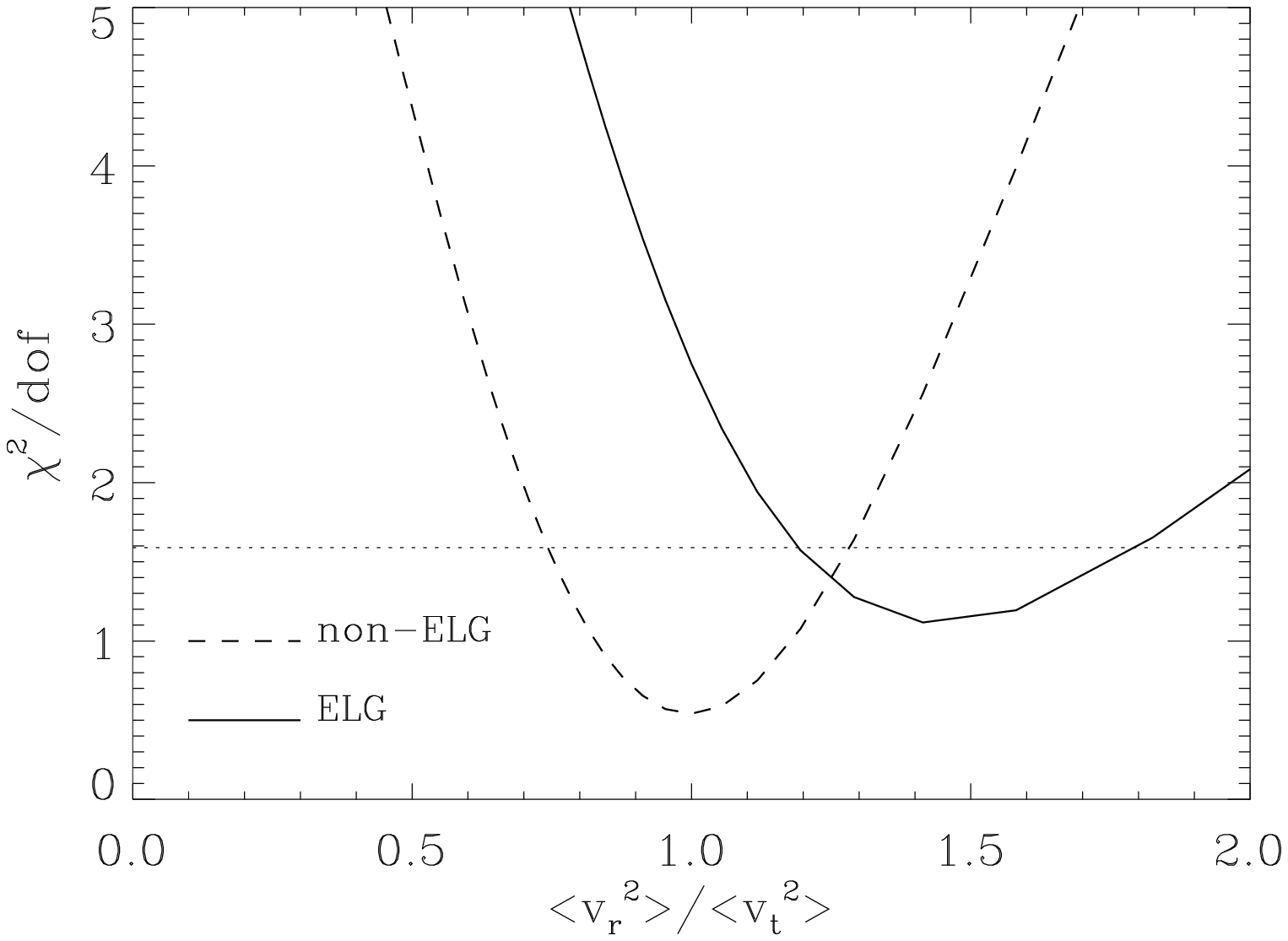}
\caption{Left panel: The velocity distribution of the non-ELG
and its decomposition into Gauss-Hermite moments.
Right panel: The $\chi^2$ solutions to the constant anisotropy
model for ELG in the gravitational potential of non-ELG (van der
Marel et al.'s method). The dotted line represents the cutoff
value above which solutions are excluded at $>99$\% confidence level.
The solutions for non-ELG in the same
potential are shown as a check for consistency.}
\end{figure}

Ever since Zwicky (1937) we have been aware of the need for estimating
cluster masses by different means. Different methods are affected by
different biases, which are sometimes difficult to properly correct
for. The perspective for improving upon the current situation is
therefore clear. We need to constrain the cluster internal dynamics by
the complementary analyses of the distribution of cluster members, as
well as of X-ray and gravitational lensing data. For this, coordinated
multi-wavelength surveys of galaxy clusters, using several
instruments, are needed. Most promising, in this respect, is the
upcoming XMM Large Scale Structure
Survey\footnote{http://vela.astro.ulg.ac.be/themes/spatial/xmm/LSS/index\_e.html},
but a much more intensive spectroscopic follow-up than currently
planned would be required to properly address the issue of internal
cluster dynamics.

\acknowledgments 
I wish to thank Stefano Borgani, Marino Mezzetti and Riccardo Valdarnini
for organizing such an interesting and entertaining workshop.
I acknowledge useful discussions with Peter Katgert, 
Marisa Girardi and Alain Mazure.

This paper is dedicated to the memory of my parents, Angelo and Anita.

\end{document}